\documentclass{JHEP3}
\usepackage{epsfig,multicol,bbm}
\usepackage{makeidx}
\usepackage{amsmath}
\usepackage{amsfonts}
\usepackage{amssymb}
\usepackage{graphicx}
\usepackage{accents}
\input{tcilatex}

\newcommand\fverb{\setbox\fverbbox=\hbox\bgroup\verb}
\newcommand\fverbdo{\egroup\medskip\noindent%
            \fbox{\unhbox\fverbbox}\ }
\newcommand\fverbit{\egroup\item[\fbox{\unhbox\fverbbox}]}
\newbox\fverbbox

\title{\textbf{ \textbf{On Black Attractors in \emph{8D }and } Heterotic/Type IIA Duality}}

\author{El Hassan Saidi\\
  \small {1. Centre of Physics and Mathematics, CPM-CNESTEN, Rabat,
Morocco,} \\
 \small {2. Laboratory of High Energy Physics, Modeling and Simulation, Faculty of Science, Rabat, Morocco},\\  E-mail: \email{
   h-saidi@fsr.ac.ma}}

 \preprint{CPM-10-01}

\abstract{Motivated by the study of black attractors in \emph{8D}
supergravity with \emph{16} supersymmetries, we use the field theory
approach and \emph{8D} supersymmetry with non trivial central
charges to shed light on the exact duality between heterotic string
on T$^{2}$ and type IIA on real connected and compact surfaces
$\Sigma _{2}$. We investigate the two constraints that should be
obeyed by $\Sigma _{2}$ and give their solutions in terms of
intersecting 2-cycles as well their classification using Dynkin
diagrams of affine Kac-Moody algebras. It is shown as well that the
moduli space of these dual theories is given by $SO\left( 1,1\right)
\times \frac{SO\left( 2,r+2\right) }{SO\left( 2\right) \times
SO\left( r+2\right) }$ where $r$ stands for the rank of the gauge
symmetry $G_{r}$ of the \emph{10D} heterotic string on $T^{2}$. The
remarkable cases $r=-2,-1,0$ as well as other features are also
investigated.}


\begin{document}


\section{Introduction}

In the last few years, there has been a growing interest in the
study the attractor mechanism of black brane configurations in
higher dimensional supergravity theories with typical moduli space
given by the coset group manifold G/H \textrm{\cite{1A}-\cite{1I}.}
This mechanism was first
discovered in the context of supersymmetric black\textrm{\ }holes in \emph{4D%
} $\mathcal{N}=2$ supergravity coupled to vector supermultiplets \textrm{%
\cite{2A}-\cite{2D}}; and was extended for black branes living in
diverse dimensions; for reviews see \textrm{\cite{3A}-\cite{3F} }and
refs therein. The attractor mechanism tells that the values $\phi
^{I}\left( r_{H}\right) $ of the moduli at the horizon geometry of
the black attractor are independent of the asymptotic values $\phi
^{I}\left( r\rightarrow \infty \right) $; and, up on solving the
attractor equations, they are completely determined by the electric
$Q_{\Lambda }$ and magnetic $P^{\Lambda }$ charges respectively
given by the fluxes $\int_{S^{6-p}}\mathcal{\tilde{F}}%
_{6-p|\Lambda }$  and $\int_{S^{p+2}}\mathcal{F}_{p+2}^{\Lambda }$
of the various massless abelian gauge fields of the supergravity
theory including p-branes.
\newline A remarkable class of the supergravity theories that have
attracted much interest in the recent literature concerns the
ultra-violet (UV) finite ones arising from superstrings and M-theory
compactifications having interpretation in terms of D- and M- branes
wrapping cycles of the internal space. There, the Maxwell type gauge
fields are associated with semi- realistic string models with
multitude of $U\left( 1\right) $ gauge factors that survive in the
typical gauge symmetry breaking
\begin{equation}
\begin{tabular}{llll}
$G_{r}$ & $\longrightarrow $ & $\dprod\limits_{i}G_{N_{i}}$ & ,%
\end{tabular}%
\end{equation}%
where $G_{r}$ can be thought of as the $E_{8}\times E_{8}$ or
$SO\left( 32\right) $ gauge symmetries of the \emph{10D} heterotic
string, or a subsymmetry of them. The $G_{N_{i}}$s are rank $N_{i}$
subgroups of $G_{r}$ with the property $\sum_{i}N_{i}\leq r\leq 16$
and associated 2-form gauge
field strengths $\mathcal{F}_{2}$ having non trivial fluxes along the $%
U^{N_{i}}\left( 1\right) $ directions and playing a central role in
the study of black attractors in these supergravity theories. In
this issue, several results have been obtained on attractor
mechanism of extremal (vanishing temperature for non-zero entropy)
BPS and non BPS black branes
living in \emph{4D}, \emph{5D}, \emph{6D}, \emph{7D} and \emph{8D }%
supergravities with various numbers of conserved supersymmetries \textrm{%
\cite{1H,1I} }and refs therein. \newline On the other hand, in
completing missing results on intersecting attractors
in non chiral $\mathcal{N}=1$ supergravity in eight dimensions\emph{\ }%
\textrm{\cite{5A} }arising from\textrm{\ \emph{10D} }heterotic
string on 2-torus and involving $\left( 2+2+n\right) $\ 1-form gauge
fields together
with $2\left( 2+n\right) $\ real scalars parameterizing the moduli space%
\textrm{\ }%
\begin{equation}
\begin{tabular}{llll}
$\frac{SO\left( 2,2+n\right) }{SO\left( 2\right) \times SO\left( 2+n\right) }%
\times SO\left( 1,1\right) $ & , & $0\leq n\leq 16$ & ,%
\end{tabular}
\label{md}
\end{equation}%
we have noticed an interesting property concerning their type IIA
interpretation using D- branes. This property, which is investigated
in this paper, concerns the two following things:

\ \ \ \ \newline (\textbf{1}) the geometric engineering of the
\emph{exact} duality between
the \emph{10D} heterotic superstring on $T^{2}$ with some gauge symmetry $%
G_{r}$; and the type IIA superstring compactified on real compact surfaces $%
\Sigma _{2}$ with D- branes wrapping cycles of $\Sigma _{2}$.

\ \ \ \newline (\textbf{2}) The heterotic/ type IIA interpretation
of the \emph{singular limit} describing the coupling of the
\emph{8D} $\mathcal{N}=1$ supergravity multiplet to \emph{one}
Maxwell gauge supermultiplet. The moduli space of this special
supersymmetric gauge theory is given by the particular real
3-dimensional coset manifold,%
\begin{equation}
\begin{tabular}{ll}
$\frac{SO\left( 2,1\right) }{SO\left( 2\right) }\times SO\left(
1,1\right) $
& ,%
\end{tabular}
\label{11}
\end{equation}%
which does not appear in the moduli space family (\ref{md});
although it
could be recovered by going beyond the lower bound of the positive integer $%
n $ and setting $n=-1$ in eq(\ref{md}).

\ \ \ \ \newline Moreover, because of the heterotic/type IIA duality
in \emph{8D}, we also show that the real compact surfaces $\Sigma
_{2}$, used in the compactification scenario
\begin{equation}
\begin{tabular}{ll}
$\mathcal{M}^{1,9}\longrightarrow \mathcal{M}^{1,7}\times \Sigma _{2}$ & ,%
\end{tabular}%
\end{equation}%
should obey two basic consistency constraints. The first one, which
is familiar in string compactification, deals with the number of
conserved supersymmetries after space time dimension reduction. The
second one however concerns the connected 2-cycle homology of
$\Sigma _{2}$ that requires a vanishing self intersection; i.e :
$\Sigma _{2}\cdot \Sigma _{2}=0$. This constraint turns out to
capture two remarkable issues:

\ \ \newline
(\textbf{a}) it allows to explain the stringy origin of the non chiral \emph{%
8D }$\mathcal{N}=1$ supergravity models with moduli spaces
$\frac{SO\left( 2,1\right) }{SO\left( 2\right) }\times SO\left(
1,1\right) $ and $SO\left( 1,1\right) $ which miss in the listing
(\ref{md}) and respectively recovered by roughly setting $n=-1,-2$.

\ \ \ \newline (\textbf{b}) it captures data on superconformal
invariance in \emph{8D} that underly known results on \emph{4D}
$\mathcal{N}=2$ conformal models. Recall that \emph{4D}
supersymmetric conformal models were first obtained in
\textrm{\cite{6A}, see also \cite{6B} for an extension, }from
\emph{10D} type II superstrings on Calabi-Yau threefolds with affine
ADE singularities; they may be re-derived from non chiral \emph{8D
}$\mathcal{N}=1$ supergravity by a further compactification on K3
complex surfaces that
reduce the \emph{16} conserved supersymmetries down to \emph{8 }\textrm{\cite%
{7A,7B,7C}}.

\ \ \ \newline To fix the ideas, let us briefly describe the two
constraints that should be obeyed by the real compact surface
$\Sigma _{2}$ used in type IIA compactification:

\begin{itemize}
\item it should preserve only \emph{16} supercharges amongst the initial
\emph{32} existing supersymmetries living in \emph{10D }type IIA
superstring. A far as duality is concerned, this implies, amongst
others, that heterotic string on $T^{2}$ and type IIA string on
$\Sigma _{2}$ should have the same underlying non chiral \emph{8D
}$\mathcal{N}=1$ supersymmetric algebra, the same types of central
charges and the same extremal black attractors.

\item the 2-cycle homology of $\Sigma _{2}$ should lead to the same fields
content one obtains in heterotic string on $T^{2}$ with some gauge
invariance $G_{r}$. This means that, on both sides, we should have
the same number of supersymmetric representations (supergravity and
Yang-Mills multiplets). In particular, we should have the same
number of \emph{8D} gauge fields $\left\{ \mathcal{A}_{\mu
}^{\Lambda }\right\} $ and the same number of \emph{8D} scalars
$\left\{ \phi ^{I}\right\} $ that parameterize the moduli space
(\ref{md}) of these theories.
\end{itemize}

\ \ \ \newline The presentation is as follows: \textrm{In section
2}, we give general features as well as constraints on non chiral
\emph{8D}, $\mathcal{N}=1$
supergravity arising from superstrings compactification. \textrm{In section 3%
}, we revisit the duality between heterotic string on T$^{2}$ and
type IIA on $\Sigma _{2}$ and show that exact duality between the
massless spectrum of the two string theories requires, amongst
others, the vanishing of the self intersection of the compact
surface; i.e: $\Sigma _{2}\cdot \Sigma
_{2}=0$. \textrm{In section 4}, we give three examples of non chiral \emph{8D%
} $\mathcal{N}=1$ supergravity models; the first one aims to
illustrate the general result of the paper; and the two others
correspond to two extreme situations including the case where the
2-torus degenerate. \textrm{In section 5}, we give our conclusion.

\section{Non chiral \emph{8D} $\mathcal{N}=1$ supergravity}

We begin by recalling that in eight space time dimensions; one has
two kinds of non chiral supergravity theories: the maximal one
arising from M-theory on 3-torus; it has \emph{32} supersymmetries
and a moduli space,
\begin{equation}
\begin{tabular}{ll}
$\boldsymbol{M}_{\text{{\small M-th/T}}^{3}}^{\emph{8D,}\mathcal{N}=2}=\frac{%
SL\left( 3,R\right) }{SO\left( 3\right) }\times \frac{SL\left( 2,R\right) }{%
SO\left( 2\right) }$ & .%
\end{tabular}%
\end{equation}%
The non chiral \emph{8D} $\mathcal{N}=1$ supergravity with a moduli space (%
\ref{md}); it may be viewed as describing the physics at Plank
energy following from two dual superstring models
\textrm{\cite{7A}}; either from the \emph{10D} heterotic superstring
compactified on $T^{2}$ with some $G_{r} $ invariance; or from
\emph{10D} type IIA superstring on some compact surface $\Sigma
_{2}$ together with D- branes wrapping cycles of $\Sigma _{2} $.
Recall that in \emph{10D} space time, the bosonic massless spectrum
of
these two theories is as follows%
\begin{eqnarray}
&&%
\begin{tabular}{llllll}
10D heterotic & : & $\mathcal{G}_{MN},$ $\mathcal{B}_{MN},$ $\Phi $ & $;$ & $%
\left\{ \mathcal{A}_{M}^{a}\right\} $ &
\end{tabular}
\label{h} \\
&&%
\begin{tabular}{llllll}
10D type IIA & : & $\mathcal{G}_{MN},$ $\mathcal{B}_{MN},$ $\Phi $ & $;$ & $%
\mathcal{A}_{M},$ $\mathcal{C}_{MNP}$ &
\end{tabular}
\label{T2}
\end{eqnarray}%
where $\mathcal{G}_{MN},$ $\mathcal{B}_{MN},$ $\Phi $ are
respectively the graviton, the NS-NS antisymmetric field and the
dilaton; and where the gauge
fields $\mathcal{A}_{M}^{a}$ are given by the trace of $\left( T^{a}\mathcal{%
A}_{M}\right) $ with $\left\{ T^{a}\right\} $ generating the $G_{r}$
invariance contained in $E_{8}\times E_{8}$ or $SO\left( 32\right) $
gauge symmetry. The other gauge fields $\mathcal{A}_{M},$
$\mathcal{C}_{MNP}$ are associated with the D0- and D2- branes of
type IIA superstring.

\subsection{Compactification to \emph{8D}}

Under compactification, the resulting massless spectrum following
from heterotic superstring on $T^{2}$ is familiar from the view of
standard quantum field theory and is easily derived. However, the
field content following from type IIA string on $\Sigma _{2}$, and
which should be the same as in the case of heterotic on T$^{2}$,
needs a careful treatment as it implies D-branes wrapping cycles of
the compact manifold. From group theoretical view, these extended
objects are manifested through a generalized supersymmetric algebra
going beyond the usual Haag-Lopozanski-Sohnius one. The latter has
central charges $Z$ carrying no Lorentz space time charges and
predicts only black holes. The generalized superalgebra involves
however operators $Z_{\mu _{1}...\mu _{p}}$ with non trivial
$SO\left( 1,7\right) $ charges having interpretation in terms of
p-branes; see eqs(\ref{sp}) for details.

\subsubsection{Deriving the homology of $\Sigma _{2}$}

Focussing on non chiral \emph{8D }supergravity with \emph{16}
supercharges,
one may use the duality in \emph{6D} space time between heterotic string on $%
T^{4}=T^{2}\times T^{2}$; and the type IIA string on $K3$ in order
to study the geometric structure of the real compact and connected
$\Sigma _{2}$ surfaces that are needed for the brane interpretation
of the extremal BPS and non BPS black attractors in \emph{8D}. By
using the adiabatic argument between the non chiral supergravities
with \emph{16} supersymmetries in \emph{6D} and \emph{8D }arising
from superstring compactifications; and roughly thinking about the
4-dimensional compact manifold (complex surface) K3 as given by the
local description $T^{2}\times S^{2}$; one ends with the following
possibility
\begin{equation}
\Sigma _{2}=S^{2}.  \label{ss}
\end{equation}%
This is a simple solution that agrees with the well known result
that the compactification on the 2-sphere preserves half of the
initial supersymmetries namely \emph{16}. But eq(\ref{ss}) let also
understand that generally speaking the compact surface $\Sigma _{2}$
may be also thought of as given by the intersection of several
copies of 2-spheres as given below
\begin{equation}
\begin{tabular}{llll}
$\Sigma _{2}=\dbigcup\limits_{a=1}^{N}\text{ }u_{a}\left[
S_{a}^{2}\right] $
& , & $u_{a}\in \mathbb{Z}_{+}$ & ,%
\end{tabular}
\label{sr}
\end{equation}%
where $u_{a}$s are positive integers that will be interpreted later
on as Dynkin weights of affine Kac-Moody algebras. As we will see,
the duality between heterotic on T$^{2}$ and type IIA on $\Sigma
_{2}$ puts a strong constraint on the intersection matrix between
the 2-spheres $S_{a}^{2}$ of the reducible 2-cycle (\ref{sr}). It
happens that $\Sigma _{2}$ should have a vanishing self intersection
\begin{equation}
\begin{tabular}{llll}
$\Sigma _{2}\cdot \Sigma _{2}$ & $=0=$ & $\dsum\limits_{a,b}u_{a}\mathcal{I}%
_{ab}u_{b}$ & $,$%
\end{tabular}
\label{SI}
\end{equation}%
with intersection matrix $\mathcal{I}_{ab}=\left[ S_{a}^{2}\right] \cdot %
\left[ S_{b}^{2}\right] .$\newline Below, we first focus on the case
(\ref{ss}) and determine the gauge fields as well as the moduli
space associated with type IIA superstring on $S^{2}$. The
investigation of the general possibility (\ref{sr}) will be done in
next section. The reason for splitting this study into two cases:
$N=1$ and $N>1,$ comes from the fact that the first case corresponds
to a singular limit which deserves to be treated separately.

\subsubsection{Type IIA string on $S^{2}$}

Starting from the spectrum (\ref{T2}) and computing the fields
content that follow from the compactification of the \emph{10D} IIA
supergravity multiplet on $S^{2}$, we get various 8D fields and too
particularly the following real 3- dimensional moduli space
\begin{equation}
\begin{tabular}{ll}
$M_{8D\text{-type IIA}}^{\mathcal{N}=1}=SO\left( 1,1\right) \times \frac{%
SO\left( 2,1\right) }{SO\left( 2\right) }$ & .%
\end{tabular}
\label{1}
\end{equation}%
Here the factor $SO\left( 1,1\right) $ is associated with the
\emph{8D} dilaton $\sigma $ while the 2- dimensional coset group
$SO\left( 2,1\right) /SO\left( 2\right) $ has to do with S$^{2}$.
The two real moduli $\left( \varphi _{1},\varphi _{2}\right) $
parameterizing this non compact manifold
correspond to the Kahler parameter of the 2-sphere and the value of the B$%
_{NS}$ field on $S^{2}$.%
\begin{equation}
\begin{tabular}{llll}
$\varphi _{1}=\int_{S^{2}}J_{\text{Kahler}}$ & , & $\varphi
_{2}=\int_{S^{2}}B_{NS}$ & .%
\end{tabular}%
\end{equation}%
Actually this is a remarkable result; since heterotic/type IIA
duality predicts at least \emph{four} scalars; that is two Maxwell
multiplets. To shed light on this property, we study here below the
moduli space of the heterotic string on a regular $T^{2}$.

\subsubsection{Heterotic string on $T^{2}$}

Starting from (\ref{h}) and doing the same thing by using the dual
description, given by the heterotic superstring on $T^{2}$ with abelian $%
U^{r}\left( 1\right) $ gauge subsymmetry, we find that the moduli
space of the resulting non chiral \emph{8D}, $\mathcal{N}=1$
supergravity is generally given by
\begin{equation}
\begin{tabular}{lll}
$SO\left( 1,1\right) \times \frac{SO\left( 2,r+2\right) }{SO\left(
2\right)
\times SO\left( r+2\right) }$ & , & $0\leq r\leq 16$%
\end{tabular}%
\end{equation}%
Here $SO\left( 1,1\right) $ is as in eq(\ref{1}) and the integer $r$
stands for the number of Maxwell gauge multiplets in the non chiral
\emph{8D} supergravity that results from the compactification of the
\emph{10D} SYM sector of the heterotic string; in other words it is
the rank of the gauge symmetry group
\begin{equation}
\begin{tabular}{llll}
$G_{r}\subseteq E_{8}\times E_{8}$ & $or$ & $SO\left( 32\right) $ & .%
\end{tabular}%
\end{equation}%
Notice that for the case $r=0$; that is in the case of pure \emph{10D }$%
\mathcal{N}\mathbf{=}1$\emph{\ }supergravity on $T^{2}$ and SYM
sector ignored, the above moduli space reduces to
\begin{equation}
\begin{tabular}{ll}
$M_{8D\text{-het}}^{\mathcal{N}=1}=SO\left( 1,1\right) \times
\frac{SO\left(
2,2\right) }{SO\left( 2\right) \times SO\left( 2\right) }$ & ,%
\end{tabular}
\label{3}
\end{equation}%
where the factor $\frac{SO\left( 2,2\right) }{SO\left( 2\right)
\times SO\left( 2\right) }$ is a\ 4-dimensional non compact manifold
parameterized by \emph{4} real moduli $\phi _{ij}$ with the
following interpretation:

\begin{itemize}
\item \emph{3} of the four real scalars namely those given by the symmetric
term $\phi _{\left( ij\right) }$ are given by the value of the metric $%
\mathcal{G}_{ij}$ on $T^{2}$; they describe the Kahler and the
complex structure of the 2-torus. For later use, we denote by
R$_{1}$ and R$_{2}$ the radii of the two 1-cycles of T$^{2}$ and set
\begin{equation}
\begin{tabular}{llll}
$k\sim R_{1}R_{2}$ & , & $\tau \sim \frac{R_{1}}{R_{2}}e^{i\vartheta }$ & .%
\end{tabular}%
\end{equation}%
respectively describing the Kahler and the shape (complex structure)
parameters of the 2-torus.

\item the fourth parameter $\phi _{\left[ ij\right] }=\varepsilon _{\left[ ij%
\right] }b$ is given by the value of the $B_{NS}$ field on $T^{2},$ i.e: $%
b=\int_{T^{2}}B_{NS}$. This modulus combine with $k\sim R_{1}R_{2}$
to give the complex Kahler modulus $t=k+ib$.
\end{itemize}

\ \newline Under compactification of the \emph{10D} supergravity
multiplet (\ref{h}) on
$T^{2}$; we also have \emph{4} gauge fields $\mathcal{G}_{\mu }^{i}$ and $%
\mathcal{B}_{\mu }^{i}$ that respectively follow from the metric $\mathcal{G}%
_{MN}$ and the NS-NS $\mathcal{B}_{MN}$ fields; two of these gauge
fields;
say $\mathcal{G}_{\mu }^{i}$, are involved in the non chiral \emph{8D} $%
\mathcal{N}=1$ supergravity multiplet while the two $\mathcal{B}_{\mu }^{i}$%
s combine with $\phi ^{ij}$ to make two \emph{8D} $\mathcal{N}=1$
Maxwell
multiplets.%
\begin{equation}
\begin{tabular}{llll}
\emph{8D} gravity & : & $\mathcal{G}_{\mu \nu },$ $\mathcal{B}_{\mu \nu },$ $%
\sigma ,$ $\mathcal{G}_{\mu }^{i}$ & , \\
\emph{8D} Maxwell & : & $\mathcal{B}_{\mu }^{i},$ $\phi ^{ij}$ & ,%
\end{tabular}
\label{gr}
\end{equation}%
with internal indices taking the values $i,j=1,2$.

\subsection{Singular limits}

By comparing the moduli spaces $M_{8D\text{-type IIA}}^{\mathcal{N}=1}$ (\ref%
{1}) and $M_{8D\text{-het}}^{\mathcal{N}=1}$ (\ref{3}), it follows
that the non chiral \emph{8D} $\mathcal{N}=1$ supergravities
following from the type IIA superstring on $S^{2}$ and the (gravity
sector of the) heterotic superstring on $T^{2}$ are not exactly
identical. As such, one concludes
that they are not exactly dual%
\begin{equation}
\begin{tabular}{llll}
type IIA on S$^{2}$ & $\nleftrightarrow $ & heterotic on $T^{2}$ & ;%
\end{tabular}
\label{t}
\end{equation}%
but still have:

\begin{itemize}
\item the same gravity supermultiplet with the bosonic fields $\mathcal{G}%
_{\mu \nu },$ $\mathcal{B}_{\mu \nu },$ $\sigma ,$ $\mathcal{G}_{\mu
}^{i}$,

\item the same non chiral \emph{8D} $\mathcal{N}=1$ superalgebra generated
by the fermionic generators $Q_{\alpha }$ and
$\bar{Q}_{\dot{\alpha}}$.
\end{itemize}

\ \ \newline Recall that, in absence of central charges, the
supercharges $Q_{\alpha }$
and $\bar{Q}_{\dot{\alpha}}$ obey the typical relation $Q_{\alpha }\bar{Q}_{%
\dot{\beta}}+\bar{Q}_{\dot{\beta}}Q_{\alpha }\sim \sigma _{\alpha \dot{\beta}%
}^{\mu }P_{\mu }$; but in general they satisfy the following
extended
anticommutation relations,%
\begin{equation}
\begin{tabular}{llll}
$\left\{ Q_{\alpha },Q_{\beta }\right\} $ & $=$ & $\delta _{\alpha
\beta
}Z_{0}+\sigma _{\left( \alpha \beta \right) }^{\mu \nu \rho \sigma }$ $%
Z_{\mu \nu \rho \sigma }$ & , \\
$\left\{ Q_{\alpha },\bar{Q}_{\dot{\beta}}\right\} $ & $=$ & $\sigma
_{\alpha \dot{\beta}}^{\mu }Z_{\mu }^{0}+\sigma _{\alpha
\dot{\beta}}^{\mu
\nu \rho }$ $Z_{\mu \nu \rho }^{0}$ & , \\
$\left\{ \bar{Q}_{\dot{\alpha}},\bar{Q}_{\dot{\beta}}\right\} $ & $=$ & $%
\delta _{\dot{\alpha}\dot{\beta}}\bar{Z}_{0}+\sigma _{\left( \alpha
\beta
\right) }^{\mu \nu \rho \sigma }$ $\tilde{Z}_{\mu \nu \rho \sigma }$ & ,%
\end{tabular}
\label{sp}
\end{equation}%
where the bosonic operators $Z_{\mu _{1}...\mu _{p}}$ are "central
charges" carrying non trivial $SO\left( 1,7\right) $ charges and
having
interpretation in terms of electric and magnetic charges of p-branes. The $%
\sigma ^{\mu _{1}...\mu _{p}}$'s are \emph{antisymmetric} products of the $%
8\times 8$ Pauli- \textrm{Dirac} $\Gamma ^{\mu }$-\ matrices in
\emph{8D}.

\ \ \ \ \ \newline Notice also that to get the exact duality
\begin{equation}
\begin{tabular}{llll}
type IIA on S$^{2}$ & $\longleftrightarrow $ & heterotic on $T^{2}$
&
\end{tabular}%
\end{equation}%
one has to put \emph{2} constraint relations on the \emph{4} real fields $%
\phi _{ij}$ ( 2 complex ones $t$ and $\tau $) in order to equate the
moduli spaces (\ref{1}) and (\ref{3}). This may be achieved either
by fixing the complex structure $\tau $ of the 2-torus or its
complexified Kahler parameter. In the last case; this corresponds to
a singular limit where one (or both) of the two circles of T$^{2}$
shrinks to zero,
\begin{equation}
\begin{tabular}{llll}
$t=k+ib$ & $\rightarrow $ & $0$ & ,%
\end{tabular}%
\end{equation}%
and happens whenever one of the two radii is reduced to zero; i.e $%
R_{i}\rightarrow 0$ in eq(\ref{t}). The singular case where both $%
t\rightarrow 0$ and $\tau \rightarrow 0$ is associated, on the side
of type IIA compactification side, with the shrinking of $S^{2}$
down to zero. This special situation corresponds precisely to
dealing with pure non chiral \emph{8D} $\mathcal{N}=1$ supergravity
with moduli space $SO\left( 1,1\right) $.

\section{Heterotic/type IIA duality in \emph{8D} revisited}

We start by recalling the degrees of freedom of non chiral \emph{8D} $%
\mathcal{N}=1$ supergravity by focussing on those used in the study
of black attractors namely the abelian gauge fields and the scalars.
Then we derive the constraint relation (\ref{SI}) on the compact
surfaces $\Sigma _{2}$ that lead to the exact heterotic/type IIA
duality in \emph{8D}. \newline Restricting the investigation to
bosons, it is interesting to notice that
the fields of the non chiral \emph{8D} $\mathcal{N}=1$ supergravity\emph{\ }%
organize into two basic supermultiplets:

\begin{itemize}
\item \emph{the gravity multiplet}: it contains, in addition to the
gravitino and graviphotino capturing \emph{40+8} on shell degrees of
freedom, the usual graviton $\mathcal{G}_{\mu \nu },$ one antisymmetric $%
\mathcal{B}_{\mu \nu }$ field, \emph{2} graviphotons denoted as $\mathcal{G}%
_{\mu }^{i}=\left( \mathcal{G}_{\mu }^{1},\mathcal{G}_{\mu
}^{2}\right) $ and the dilaton $\sigma $. The on shell degrees of
freedom of these fields
are respectively as follows:%
\begin{equation}
20+15+2\times 6+1=48
\end{equation}

\item N \emph{Maxwell multiplets}: they contain $N$ gauge fields $\mathcal{A}%
_{\mu }^{a}=\left( \mathcal{A}_{\mu }^{1},...,\mathcal{A}_{\mu
}^{N}\right) $ and $2N$ real scalar fields $\phi ^{ia}=\left( \phi
^{i1},...,\phi ^{iN}\right) $ and $i=1,2$. The number of gauge
supermultiplets depends on
the gauge symmetry of the theory. In the case of non chiral \emph{8D} $%
\mathcal{N}=1$ supergravity arising from heterotic string on regular
2-torus, this number should be as%
\begin{equation}
N\leq 18.
\end{equation}
\end{itemize}

\ \ \ \ \ \ \ \ \ \ \ \ \ \ \newline The gauge fields
$\mathcal{G}_{\mu }^{i}$ and $\mathcal{A}_{\mu }^{a}$ as well as the
real scalar moduli $\phi ^{ia}$ have geometric and stringy meaning
in the framework of superstring compactifications; but with
different interpretations depending on whether they are following
from heterotic string on T$^{2}$ or from type IIA string on $\Sigma
_{2}$.

\subsection{Heterotic string on T$^{2}$}

In the case of heterotic string on T$^{2}$ with a $U^{r}\left(
1\right) $ abelian gauge subsymmetry, the $2N$ real scalars $\phi
^{ia}$ of the non chiral \emph{8D} $\mathcal{N}=1$ supergravity
decomposes like
\begin{equation}
\begin{tabular}{llll}
$2N=4+2r$ & , & $N\geq 0$ &
\end{tabular}%
\end{equation}%
and are interpreted as follows:

\begin{itemize}
\item \emph{the real 4 scalars}: they split as \emph{3+1 }having respectively%
\emph{\ }a geometric and stringy origin; the three scalars describe
the
Kahler and complex structure of the 2-torus; the fourth is given by the B$%
_{NS}$- field on T$^{2}$. These are precisely the complex fields
moduli $t$ and $\tau $ encountered previously.

\item $2r$\emph{\ real scalars} $\phi ^{ia}$: they come from the $%
U^{r}\left( 1\right) $ gauge fields of the SYM$_{10}$ sector of the
heterotic string along the 8-th and 9-th compact directions as given
below
\begin{equation}
\begin{tabular}{llll}
$\phi ^{ia}=\left(
\begin{array}{c}
\mathcal{A}_{8}^{a} \\
\mathcal{A}_{9}^{a}%
\end{array}%
\right) $ & , & $a=1,...,r$ & .%
\end{tabular}%
\end{equation}%
Notice that $U^{r}\left( 1\right) $ is just the Cartan subgroup of
the gauge symmetry $G_{r}\subseteq E_{8}\times E_{8}$ or $SO\left(
32\right) $. Notice also that the $T^{2}$ directions fill the space
coordinates $\left( x^{8},x^{9}\right) $.
\end{itemize}

\subsection{Type IIA superstring on $\Sigma _{2}$}

Here we use a field theoretical method and requires the \emph{exact}
heterotic/type IIA duality to derive the 2-cycle homology of the
compact surfaces $\Sigma _{2}$ as well as the constraint relation
(\ref{SI}).

\subsubsection{Field theory method}

In non chiral \emph{8D} $\mathcal{N}=1$ supergravity with a generic
moduli space $\frac{SO\left( 2,N\right) }{SO\left( 2\right) \times
SO\left( N\right) }\times SO\left( 1,1\right) $, the 1-form gauge
fields and the scalars are as follows:

\begin{itemize}
\item \emph{1-form} \emph{gauge fields}: there are $\left( 2+N\right) $
gauge fields that may be written collectively like
\begin{equation}
\begin{tabular}{ll}
$\mathcal{A}_{\mu }^{\Lambda }=\left( \mathcal{G}_{\mu }^{i},\mathcal{C}%
_{\mu }^{a}\right) $ & ,%
\end{tabular}%
\end{equation}%
where for later use, it is useful to think about the two graviphotons $%
\mathcal{G}_{\mu }^{i}$ as follows
\begin{equation}
\begin{tabular}{ll}
$\mathcal{G}_{\mu }^{i}=\left( \mathcal{A}_{\mu
}^{0},\mathcal{C}_{\mu
}^{0}\right) $ & .%
\end{tabular}
\label{G}
\end{equation}%
These $\left( 2+N\right) $ gauge fields transform differently under the $%
SO\left( 2\right) \times SO\left( N\right) $ symmetry of the moduli
space. We have:
\begin{equation}
\begin{tabular}{llllllll}
$\mathcal{G}_{\mu }^{i}$ & $\sim $ & $\left( 2,1\right) $ & , & $\mathcal{C}%
_{\mu }^{a}$ & $\sim $ & $\left( 1,N\right) $ & .%
\end{tabular}
\label{GC}
\end{equation}%
In the type IIA picture on $\Sigma _{2}$, the above gauge fields are
associated with two kinds of D0- branes as described below:

\begin{itemize}
\item the gauge field $\mathcal{A}_{\mu }^{0}$ of eq(\ref{G}); it is
associated with the standard D0- brane that descend from \emph{10D}
type IIA.

\item the gauge fields $\mathcal{C}_{\mu }^{0}$ of eq(\ref{G}) and the $N$
gauge fields $\mathcal{C}_{\mu }^{a}$ of eq(\ref{GC}); they are
associated with those D0-branes following from D2-branes wrapping
2-cycles in $\Sigma _{2}$.
\end{itemize}
\end{itemize}

\ \ \ \ \newline From the second property as well as the fact that
half of the \emph{32}
supersymmetries should be preserved, we learn that the compact surface $%
\Sigma _{2}$ should have $\left( 1+N\right) $ irreducible 2-spheres
as shown
below.%
\begin{equation}
\begin{tabular}{llll}
$\Sigma _{2}^{\left( N+1\right) }=\dbigcup\limits_{a=0}^{N}$
$u_{a}\left[
S_{a}^{2}\right] $ & , & $u_{a}\in \mathbb{Z}_{+}$ & .%
\end{tabular}
\label{SN}
\end{equation}%
where the positive integers $u_{a}$ are as mentioned before.
\newline Using this 2-cycle homology, one clearly sees that the
gauge fields are
given by%
\begin{equation}
\begin{tabular}{llll}
$\mathcal{C}_{\mu }^{a}=\int_{S_{a}^{2}}\mathcal{C}_{\mu z\bar{z}}$ $%
dz\wedge d\bar{z}$ & , & $a=0,1,...,N$ & ,%
\end{tabular}%
\end{equation}%
where $z_{a},$ $\bar{z}_{a}$ are complex coordinates that
parameterize the 2-sphere $S_{a}^{2}$ and $\mathcal{C}_{\mu
z\bar{z}}$ is the gauge 3- form associated with D2- brane wrapping
$S_{a}^{2}$.\newline However to determine the intersection matrix
$\mathcal{I}_{ab}$ between the irreducible 2- spheres; we need an
extra information that turns out to be exactly given by matching the
number of the real scalar fields in the supergravity. This feature
is discussed below.

\begin{itemize}
\item \emph{matching the} \emph{scalar fields}: Besides the dilaton $\sigma $%
, there are $2N$ real scalars $\phi ^{ia}$ transforming under the
$SO\left( 2\right) \times SO\left( N\right) $ symmetry of the moduli
space as follows,
\begin{equation}
\begin{tabular}{llll}
$\phi ^{ia}$ & $\sim $ & $\left( 2,N\right) $ & .%
\end{tabular}
\label{2N}
\end{equation}%
Since each 2-sphere contributes with two real scalars $\varphi _{1}$ and $%
\varphi _{2}$; one associated with the Kahler parameters and the
other given
by the value of the B$_{NS}$ field $S_{a}^{2}$, we have the following.%
\newline
A complex scalar given by%
\begin{equation}
\begin{tabular}{ll}
$\Phi ^{0}=\int_{S_{0}^{2}}\left( J_{\text{Kahler}}+iB_{NS}\right) $ & ,%
\end{tabular}%
\end{equation}%
and N similar others as follows:
\begin{equation}
\begin{tabular}{llll}
$\Phi ^{a}=\int_{S_{a}^{2}}\left( J_{\text{Kahler}}+iB_{NS}\right) $ & , & $%
a=1,...,N$ & ,%
\end{tabular}%
\end{equation}%
where $J_{\text{Kahler}}+iB_{NS}$ stands for the complexified Kahler
form and where we have set $\Phi =$ $\varphi _{1}+i\varphi _{2}$.
\end{itemize}

\ \ \ \newline From this construction, we learn that there are two
real (one complex) undesired scalars since $N$ Maxwell
supermultiplets need only $2N$ real scalars as in eq(\ref{2N}). This
means that the (complex) scalars $\Phi _{a}$ should obey one complex
(two real) condition. A natural constraint equation corresponds to
express $\Phi _{0}$ in terms of the others as given by the following
linear relation,
\begin{equation}
\begin{tabular}{llll}
$\Phi _{0}$ & $=$ & $\dsum\limits_{b=1}^{r}\lambda _{b}\Phi _{b}$ & ,%
\end{tabular}
\label{F}
\end{equation}%
that lead to the appropriate number of scalars. To get the explicit
expression of the coefficients $\lambda _{b}$, we use a well known
result \textrm{\cite{8A,8B}} giving a correspondence between
(irreducible) 2-spheres $S_{a}^{2}$ and (simple) roots
$\mathbf{\alpha }_{a}$ of affine
Kac-Moody algebras,%
\begin{equation}
\begin{tabular}{llll}
$\left[ S_{a}^{2}\right] $ & $\longleftrightarrow $ &
$\mathbf{\alpha }_{a}$
& , \\
&  &  &  \\
$\mathbf{\alpha }_{0}$ & $=$ & $\delta -\sum_{b=1}^{r}u_{b}\mathbf{\alpha }%
_{b}$ & .%
\end{tabular}
\label{CO}
\end{equation}%
In the second relation of the above eqs, which should be put in 1:1 with eq(%
\ref{F}), the $\mathbf{\alpha }_{0}$ is the simple root that is
associated
with affine node in the Dynkin diagram of affine Kac-Moody algebras, $%
\mathbf{\alpha }_{1},...,\mathbf{\alpha }_{r}$ are the usual simple
roots of finite dimensional Lie algebras and $u_{b}$ the Dynkin
weights. $\delta $ is an imaginary root; we will ignore it
here.\newline
Using this correspondence, we end with an intersection matrix $\mathcal{I}%
_{ab}=\left[ S_{a}^{2}\right] \cdot \left[ S_{b}^{2}\right] $ given
by minus the generalized Cartan matrix $\mathcal{K}_{ab}$ of affine
Kac-Moody
algebras. More precisely, we have%
\begin{equation}
\mathcal{I}_{ab}=-\mathcal{K}_{ab}
\end{equation}%
with%
\begin{equation}
\mathcal{K}_{ab}=\frac{2\left( \mathbf{\alpha }_{a},\mathbf{\alpha }%
_{b}\right) }{\left( \mathbf{\alpha }_{a},\mathbf{\alpha
}_{a}\right) }
\end{equation}%
With this intersection matrix and the identities $\sum_{a}u_{a}\mathcal{I}%
_{ab}=0$ and $\sum_{b}\mathcal{I}_{ab}u_{b}=0$, we end with the
following self intersection property $\Sigma _{2}\cdot \Sigma
_{2}=0$.

\subsubsection{Result}

The exact duality between heterotic string on $T^{2}$ with some
gauge symmetry $G_{r}$ and type IIA string on $\Sigma _{2}$
requires:

\begin{itemize}
\item real compact and connected surfaces type%
\begin{equation*}
\begin{tabular}{llll}
$\Sigma _{2}$ & $=$ & $\dbigcup\limits_{a=1}^{N}u_{a}\left[
S_{a}^{2}\right] $ &
\end{tabular}%
\end{equation*}%
with the properties:

\begin{itemize}
\item have vanishing self intersection $\Sigma _{2}\cdot \Sigma _{2}=0$

\item preserve \emph{16} of the \emph{32} supersymmetries
\end{itemize}

\item surfaces classified by the affine Kac-Moody extension of the $G_{r}$
since:

\begin{itemize}
\item the intersection matrix $\mathcal{I}_{ab}=\left[ S_{a}^{2}\right]
\cdot \left[ S_{b}^{2}\right] $ between 2-spheres of $\Sigma _{2}$
is given by minus the generalized Cartan matrix $K_{ij}\left(
\hat{g}\right) $ of
affine Kac-Moody algebras $\hat{g}$%
\begin{equation}
\begin{tabular}{ll}
$\mathcal{I}_{ab}=-K_{ab}\left( \hat{g}\right) $ & .%
\end{tabular}
\label{im}
\end{equation}%
\ Since affine Kac-Moody algebras $\hat{g}$ are classified; it
follows that the $\Sigma _{2}$'s are given by the Dynkin diagrams of
the affine Kac-
Moody- Lie algebras. In the case symmetric generalized Cartan matrices $%
K_{ab}=K_{ba}$, these surfaces are given by the Dynkin diagrams of
the following simply laced Kac-Moody algebras
\begin{equation}
\begin{tabular}{lllll}
$\widehat{su}\left( m\right) ,$ & $\widehat{so}\left( 2m\right) ,$ & $\hat{E}%
_{6},$ & $\hat{E}_{7},$ & $\hat{E}_{8},$%
\end{tabular}%
\end{equation}

\item A graphic representation of examples of $\Sigma _{2}\left( \hat{g}%
_{r}\right) $ are depicted in figure 1 where the 2-spheres are given
by the nodes of the quivers and the intersection by lines joining
the nodes.
\end{itemize}
\end{itemize}

\begin{figure}[tbph]
\begin{center}
\hspace{0cm} \includegraphics[width=14cm]{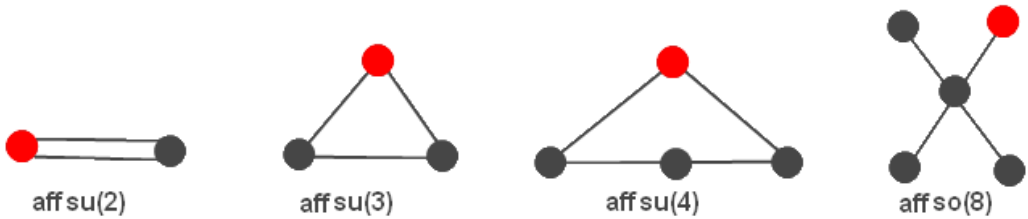}
\end{center}
\par
\vspace{-0.5cm} \caption{{\protect\small Four examples of compact
real surfaces given by intersecting 2-spheres. These quivers are in
one in to one correspondence with the affine Dynkin diagrams; the
red node gives the affine extension.}} \label{FIG1}
\end{figure}
These surfaces $\Sigma _{2}\left( \hat{g}_{r}\right) $ capture also
the nature of the non abelian gauge symmetry that may live in the
non chiral
\emph{8D} $\mathcal{N}=1$ supergravity theory. Since affine Lie algebras $%
\hat{g}_{r}$ are built out of ordinary Lie algebras $g_{r}$; one can
recover the non abelian gauge invariance by switching off the vevs
(moduli). \newline Below we discuss three examples of dual models
namely those involving
compact and connected surfaces given by the Dynkin diagrams of $\widehat{su}%
\left( 3\right) $\emph{, }$\widehat{so}\left( 32\right) $ and $\widehat{su}%
\left( 2\right) $.

\section{Heterotic/type IIA dual models}

In this section, we study three models using duality between type
IIA string a on compact real surface $\Sigma _{2}\left(
\hat{g}_{N}\right) $ and heterotic string on T$^{2}$ with
$U^{N-1}\left( 1\right) $ gauge symmetry. These are:

\begin{itemize}
\item the $\widehat{su}\left( 3\right) $\emph{\ }model, which we use to
illustrate the general idea,

\item the $\widehat{so}\left( 32\right) $ model since the group $SO\left(
32\right) $ is one of the two largest gauge symmetries of the
\emph{10D} heterotic string theory,

\item the $\widehat{su}\left( 2\right) $\emph{\ }model as it corresponds to
a singular limit for heterotic string on T$^{2}$.
\end{itemize}

\subsection{$\widehat{su}\left( 3\right) $\emph{\ }model}

In this supergravity model, the compact surface $\Sigma _{2,\widehat{su}%
_{3}} $ is associated with $\widehat{su}\left( 3\right) $ affine
Kac-Moody algebra having three simple roots $\alpha _{0},$ $\alpha
_{1}$, $\alpha _{2}$
with $\alpha _{0}=-\alpha _{1}-\alpha _{2}$. Using the correspondence (\ref%
{CO}), we then have
\begin{equation}
\begin{tabular}{lll}
$\Sigma _{2,\widehat{su}_{3}}=$ & $\left[ S_{0}^{2}\right] \dbigcup
\left[ S_{1}^{2}\right] \dbigcup \left[ S_{2}^{2}\right] $ &
\end{tabular}%
\end{equation}%
with intersection matrix like
\begin{equation}
\begin{tabular}{ll}
$\mathcal{I}_{ab}=\left(
\begin{array}{ccc}
-2 & +1 & +1 \\
+1 & -2 & +1 \\
+1 & +1 & -2%
\end{array}%
\right) $ &
\end{tabular}%
\end{equation}%
The 1-form gauge fields as well as the moduli of this model are as
follows

\begin{itemize}
\item \emph{4} Maxwell gauge fields partitioned as follows:

\begin{itemize}
\item the $\mathcal{A}_{\mu }^{0}$ graviphoton associated with the D0-brane (%
\ref{G});

\item \emph{3} Maxwell gauge fields $\mathcal{C}_{\mu }^{a}$ given by%
\begin{equation}
\begin{tabular}{lll}
$\mathcal{C}_{\mu }^{a}=\int_{S_{a}^{2}}\mathcal{C}_{\mu z\bar{z}}$ $%
dz\wedge d\bar{z}$ & , & $a=0,1,2.$%
\end{tabular}%
\end{equation}%
These \emph{3} fields correspond to D2-brane wrapping the three 2-spheres $%
S_{a}^{2}$.
\end{itemize}

\item 4 real scalars:

A priori there are \emph{3 }complex ( 6 real) moduli coming from the
values
of the complexified Kahler form on the 2-spheres of $\Sigma _{2,\widehat{su}%
_{3}}$%
\begin{equation}
\begin{tabular}{ll}
$\Phi _{a}=\int_{S_{a}^{2}}\left( J_{\text{Kahler}}+iB_{NS}\right) $ & .%
\end{tabular}%
\end{equation}

But these field are not independent; they obey the condition
\begin{equation}
\Phi _{0}+\Phi _{1}+\Phi _{2}=0
\end{equation}%
associated with Lie algebraic relation $\alpha _{0}+\alpha
_{1}+\alpha _{2}=0 $.
\end{itemize}

\ \ \newline
As such one is left with 4 real scalars parameterizing the moduli space $%
\frac{SO\left( 2,2\right) }{SO\left( 2\right) \times SO\left(
2\right) }$.
\newline
This model is dual to the \emph{10D} heterotic string on T$^{2}$
where the fluxes of the gauges fields in the SYM$_{8}$ sector and
the VEVs of the
corresponding scalars completely vanish.%
\begin{equation}
\begin{tabular}{llllll}
$\int_{S^{2}}\mathcal{F}_{2}^{\left( {\scriptsize G}_{r}\right) }=0$ & , & $%
\int_{S^{6}}\mathcal{\tilde{F}}_{6}^{\left( {\scriptsize
G}_{r}\right) }=0$
& , & $<\Phi >=0$ & ,%
\end{tabular}%
\end{equation}%
with the field strength 2-forms $\mathcal{F}_{2}^{\left( {\scriptsize G}%
_{r}\right) }=\sum \tau _{I}\mathcal{F}_{2}^{I}$, their Hodge duals $%
\mathcal{\tilde{F}}_{6}^{\left( {\scriptsize G}_{r}\right) }=\sum \tau _{I}%
\mathcal{\tilde{F}}_{6}^{I}$ and the adjoint matter $\Phi =\sum \tau
_{I}\Phi ^{I}$ are valued in the Lie algebra $G_{r}\subset
E_{8}\times E_{8}$ or $SO\left( 32\right) $ gauge symmetry with
generators $\tau _{I}$.

\subsection{$\widehat{so}\left( 32\right) $\emph{\ }model}

In this \emph{8D} supergravity model, the compact real surface $\Sigma _{2}$%
, whose quiver is depicted in figure 2, is given by the intersection
of
seventeen 2-spheres%
\begin{equation}
\begin{tabular}{lll}
$\Sigma _{2}=$ & $\dbigcup\limits_{a=0}^{16}u_{a}\left[
S_{a}^{2}\right] $ &
\end{tabular}%
\end{equation}%
with Dynkin weights $u_{a}=\left( 1,1,2,2...2,2,1,1\right) $ and
intersection matrix $\mathcal{I}_{ab}$ given by minus the
generalized Cartan
matrix $-K_{ab}^{\widehat{so}\left( 32\right) }$ of the affine Kac-Moody $%
\widehat{so}\left( 32\right) $.

\begin{figure}[tbph]
\begin{center}
\hspace{0cm} \includegraphics[width=8cm]{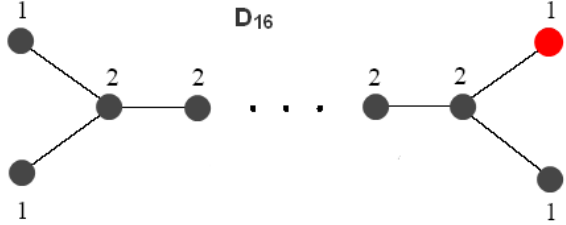}
\end{center}
\par
\vspace{-0.5cm}
\caption{{\protect\small Dynkin diagram of affine }${\protect\small SO}%
\left( {\protect\small 32}\right) $}
\end{figure}
This model involves \emph{18} Maxwell gauge fields; \emph{2} of them
(graviphotons) belong to the gravity multiplet and the \emph{16}
others are valued in the Cartan subalgebra of the $SO\left(
32\right) $ gauge symmetry.
There are also \emph{32} scalar fields parameterizing the moduli space $%
\frac{SO\left( 2,16\right) }{SO\left( 2\right) \times SO\left(
16\right) }$.

\subsection{$\widehat{su}\left( 2\right) $\emph{\ }model}

This is a simple and special model that follows from type IIA
superstring on
a compact real surface given by the intersection of two 2-spheres%
\begin{equation}
\begin{tabular}{lll}
$\Sigma _{2}=$ & $\left[ S_{0}^{2}\right] \dbigcup \left[
S_{1}^{2}\right] $
& ,%
\end{tabular}%
\end{equation}%
with intersection matrix as
\begin{equation}
\begin{tabular}{ll}
$\mathcal{I}_{ab}^{\widehat{su}_{2}}=\left(
\begin{array}{cc}
-2 & +2 \\
+2 & -2%
\end{array}%
\right) $ & .%
\end{tabular}%
\end{equation}%
The resulting non chiral \emph{8D} $\mathcal{N}=1$ supergravity has,
amongst others, \emph{3} gauge fields and one complex scalar in
addition to the dilaton $\sigma $. These fields are as follows:

\begin{itemize}
\item \emph{3} \emph{Maxwell gauge fields:} they are given by:

\begin{itemize}
\item the $\mathcal{A}_{\mu }^{0}$ graviphoton associated with the D0-brane (%
\ref{G});

\item the $\mathcal{C}_{\mu }^{0}$ graviphoton associated with the D2-brane
wrapping $S_{0}^{2}$ and reading as
\begin{equation}
\begin{tabular}{ll}
$\mathcal{C}_{\mu }^{0}=\int_{S_{0}^{2}}\mathcal{C}_{\mu z\bar{z}}$ $%
dz\wedge d\bar{z}$ & ,%
\end{tabular}%
\end{equation}%
with $\mathcal{C}_{MNP}$ the RR gauge 3-form of the \emph{10D} type
IIA spectrum.

\item the\emph{\ }Maxwell gauge fields $\mathcal{C}_{\mu }^{1}$ associated
with the D2-brane wrapping $S_{1}^{2}$; it is given by%
\begin{equation}
\begin{tabular}{ll}
$\mathcal{C}_{\mu }^{1}=\int_{S_{1}^{2}}\mathcal{C}_{\mu z\bar{z}}$ $%
dz\wedge d\bar{z}$ & ,%
\end{tabular}%
\end{equation}
\end{itemize}

\item one free complex scalar $\Phi $ (\emph{2} real scalars) solving the
condition the constraint relation $\Phi _{0}+\Phi _{1}=0$ with $\Phi
_{0}$
and $\Phi _{1}$ given by the complexified Kahler parameters%
\begin{equation}
\begin{tabular}{ll}
$\Phi _{0}=\int_{S_{0}^{2}}\left( J_{\text{Kahler}}+iB_{NS}\right) $ & ,. \\
$\Phi _{1}=\int_{S_{1}^{2}}\left( J_{\text{Kahler}}+iB_{NS}\right) $ & ,%
\end{tabular}
\label{k}
\end{equation}%
Recall that the constraint relation $\Phi _{0}+\Phi _{1}=0$ is
associated with the $\widehat{su}\left( 2\right) $ affine Kac-Moody
algebraic relation between the simple roots namely $\alpha
_{0}+\alpha _{1}=0$.
\end{itemize}

\ \ \ \ \newline This $\widehat{su}\left( 2\right) $ involves
\emph{2} real moduli rather than the \emph{4 }scalars that we have
in the \emph{10D} heterotic string on $T^{2}$ whose moduli space is
given by $\frac{SO\left( 2,2\right) }{SO\left( 2\right) \times
SO\left( 2\right) }$ as shown on eq(\ref{3}). It corresponds
therefore to a singular configuration where the complexified Kahler
structure
\begin{equation}
\begin{tabular}{llll}
$t+ib$ & $=$ & $\int_{a_{1}\times a_{2}}\left( J_{\text{Kahler}%
}+iB_{NS}\right) $ & ,%
\end{tabular}%
\end{equation}%
and the values of the the ten dimensional $B_{MN}$ field along the
two 1-cycles $a_{i}$, generating the periods of the 2-torus T$^{2}$,
\begin{equation}
\begin{tabular}{llll}
$B_{\mu i}$ & $=$ & $\int_{a_{i}}B_{\mu \nu }dy^{\nu }$ & ,%
\end{tabular}%
\end{equation}%
vanish identically. But the torus still has a complex structure
described by a complex moduli $\tau $ that should be put in one to
one correspondence with the $\Phi $ of (\ref{k}).

\section{Conclusion}

Motivated by the study of black attractors in non chiral \emph{8D} $\mathcal{%
N}=1$ supergravity with gauge 2-form $\mathcal{B}_{\mu \nu }$,
$\left( 2+N\right) $ Maxwell gauge fields $\mathcal{A}_{\mu
}^{\Lambda }=\left( \mathcal{A}_{\mu }^{i},\mathcal{A}_{\mu
}^{a}\right) $ and $\left( 1+2N\right) $ real scalars $\left( \sigma
,\phi ^{ia}\right) $ parameterizing the moduli space
\begin{equation}
SO\left( 1,1\right) \times \frac{SO\left( 2,N\right) }{SO\left(
2\right) \times SO\left( N\right) },  \label{NN}
\end{equation}%
and using an explicit field theoretic method, we have studied in
this paper
two main things: First, we have investigated the exact duality\emph{\ }%
between the heterotic string on a \emph{regular} $T^{2}$, with
$U^{r}\left( 1\right) $ abelian symmetry contained in some rank r
gauge group $G_{r}$, and the type IIA string on a connected compact
real surface $\Sigma _{2}$. Because of the fact that $\Sigma _{2}$
should preserve \emph{16} supersymmetries; that is the half of the
\emph{32} existing supercharges in the \emph{10D} type IIA
superstring and because of heterotic/type IIA duality in \emph{8D},
we have shown that connected and compact $\Sigma _{2}$
should have a vanishing self intersection,%
\begin{equation}
\Sigma _{2}\cdot \Sigma _{2}=0,
\end{equation}%
that is having the same homology as the 2-torus used in the
heterotic side. Moreover using known results on geometric
engineering of QFTs embedded in string theory, we have shown that
2-cycle homology of the surface $\Sigma _{2}$ is given by
intersecting 2-spheres $S_{a}^{2}$ with intersection
matrices $\mathcal{I}_{ab}$ classified by the generalized Cartan matrices $%
\mathcal{K}_{ab}\left( \hat{g}\right) $ of the affine Kac-Moody Lie
algebras $\hat{g}$.\newline Second, we have studied some special
dual models; in particular the non chiral \emph{8D} $\mathcal{N}=1$
supergravity theory having \emph{3} Maxwell gauge fields and
\emph{1+2} real scalar parameterizing the moduli space
\begin{equation}
SO\left( 1,1\right) \times \frac{SO\left( 2,1\right) }{SO\left(
2\right) }.
\end{equation}%
This real 3- dimensional manifold does not belong to the well known family $%
SO\left( 1,1\right) \times \frac{SO\left( 2,2+r\right) }{SO\left(
2\right) \times SO\left( 2,r\right) }$ with $0\leq r\leq 16$. We
have shown that, from heterotic side, this supergravity model
follows from compactification on a \emph{singular} torus $T^{2}$
with vanishing complexified Kahler structure. In the type IIA
picture, the model follows from the
compactification on a real surface given by two intersecting 2-spheres $%
\left[ S_{0}^{2}\right] \dbigcup \left[ S_{1}^{2}\right] $ with
branch cuts and matrix intersection
\begin{equation}
\begin{tabular}{llllll}
$\left[ S_{0}^{2}\right] \cdot \left[ S_{0}^{2}\right] =-2$ & , &
$\left[
S_{1}^{2}\right] \cdot \left[ S_{1}^{2}\right] =-2$ & , & $\left[ S_{0}^{2}%
\right] \cdot \left[ S_{1}^{2}\right] =2$ &
\end{tabular}%
\end{equation}
given by minus the generalized Cartan matrix of the twisted affine
Kac-Moody algebra $\widehat{su}\left( 2\right) $. Accordingly, one
of the \emph{3} gauge fields is associated with the D0 brane while
the 2 others follow from the D2 brane wrapping the 2-spheres of
$\Sigma _{2}$. We end this study by noting that it would be
interesting to explore the issue regarding the properties of type
IIA compactification on non connected surfaces. Progress in this
direction will reported in a future occasion.

\section{Acknowledgements}

The Centre of Physics and Mathematics (CPM), Rabat-Morocco, is a
tripartite agreement between:

\begin{itemize}
\item the Abdus Salam International Center for Theoretical Physics (ICTP),
Trieste, Italy,

\item the Ministry of Higher Education and Scientific Research (MENESFCRS),
Morocco,

\item the National Center for Nuclear Energy, Sciences and Techniques (
CNESTEN), Rabat-Morocco.
\end{itemize}

\ \ \newline CPM is a regional scientific centre hosted by CNESTEN,
Rabat-Morocco. Its main objective is to contribute to develop
training and scientific research activities in the fields of physics
and mathematics; in Morocco, North African and sub-Saharan
countries.

\ \ \ \ \newline I would like to warmly thank Professor Fernando
Quevedo, Director of ICTP; Dr. Khalid El Mediouri, Director General
of CNESTEN; Dr. Ahmed El Hattab, Director of the Division of Exact
Sciences, MENESFCRS- Morocco; Prof. Wail Benjelloun, the Dean of
Faculty of Sciences of Rabat (FSR); Prof Mohamed Lferde, Director of
CED (FSR); and Prof. Ahmed Mzerd the head of the Physics Department-
FSR, for their support to CPM.

\ \ \ \newline Special thanks to Prof. Seif Randjbar-Daemi,
Assistant of the ICTP, and Head of the Section of High Energy
Physics and Cosmology, ICTP, Trieste- Italy; who worked very hard in
the establishment of CPM.

\ \ \ \ \ \newline I would like to also thank all professors,
researchers, PhD students of Lab High Energy Physics-Modeling and
Simulations (LPHE-MS), the partners in the Moroccan Grouping in HEP
(GNPHE) and the Moroccan Network in HEP and Mathematical Physics
(RHEMAP) for their contributions and their active participation to
the ongoing CPM activities.\

\ \ \ \newline Thanks as well to the colleagues and collaborators
from the Mediterranean region and the international scientific
partners for help to CPM and their continuous support to the HEP
group of Rabat.
\newline

\ \ \ \ \ \


\begin{thebibliography}{99}
\bibitem{1A} Anna Ceresole, Sergio Ferrara,\emph{\ Black Holes and
Attractors in Supergravity}, arXiv:1009.4175,

\bibitem{1B} L. Andrianopoli, R. D'Auria, S. Ferrara, P. Fr\'{e}, M.
Trigiante, R--R Scalars, \emph{U--Duality and Solvable Lie
Algebras,} Nucl.Phys. B496 (1997) 617-629, arXiv:hep-th/9611014,

\bibitem{1C} Sergio Ferrara, Kuniko Hayakawa, Alessio Marrani, \emph{Erice
Lectures on Black Holes and Attractors},
Fortsch.Phys.56:993-1046,2008, arXiv:0805.2498,

\bibitem{1D} S. Bellucci, S. Ferrara, A. Marrani,\emph{\ Attractors in Black}%
, Fortsch.Phys.56:761-785,2008, arXiv:0805.1310,

\bibitem{1E} S. Ferrara and R. Kallosh,\emph{\ Universality of
Supersymmetric Attractors}, Phys. Rev.D54, 1525 (1996),
arXiv:9603090,

\bibitem{1F} M. Gunaydin, G. Sierra and P. K. Townsend,\emph{\ Exceptional
Supergravity Theories and the Magic Square}, Phys. Lett. B133, 72
(1983),

\bibitem{1G} E.H Saidi, A. Segui, \emph{Entropy of Pairs of Dual Attractors
in 6D/7D}, J. High Energy Phys. JHEP07(2008)128, arXiv:0803.2945,

\bibitem{1H} S. Ferrara, A. Marrani, J. F. Morales, H. Samtleben, \emph{%
Intersecting Attractor},\emph{\ }Phys.Rev.D79:065031,2009,
arXiv:0812.0050,

\bibitem{1I} L. B. Drissi, F. Z. Hassani, H. Jehjouh, E. H. Saidi, \emph{%
Extremal Black Attractors in 8D Maximal Supergravity},
PhysRevD.81.105030,2010, arXiv:1008.2689,

\bibitem{2A} S. Ferrara, R. Kallosh and A. Strominger, N = 2 Extremal Black
Holes, Phys. Rev.D52 (1995) 5412, hep-th/9508072,

\bibitem{2B} S. Ferrara and R. Kallosh, Supersymmetry and Attractors, Phys.
Rev. D54 (1996) 1514, hep-th/9602136,

\bibitem{2C} A. Strominger, Macroscopic Entropy of N = 2 Extremal Black
Holes, Phys. Lett. B383 (1996) 39, hep-th/9602111,

\bibitem{2D} Lilia Anguelova, \emph{Flux Vacua Attractors and Generalized
Compactifications}, JHEP 0901:017,2009, arXiv:0806.3820

\bibitem{3A} S. Ferrara, K. Hayakawa and A. Marrani, Erice Lectures on Black
Holes and Attractors, arXiv:0805.2498,

\bibitem{3B} S. Bellucci, S. Ferrara, R. Kallosh and A. Marrani, Extremal
Black Hole and Flux Vacua Attractors, arXiv:0711.4547,

\bibitem{3C} F. Larsen, \emph{The Attractor Mechanism in Five Dimensions},
hep-th/0608191,

\bibitem{3D} A. Belhaj, L. Drissi, E. Saidi and A. Segui, $\mathcal{N}=2$
Supersymmetric Black Attractors in Six and Seven Dimensions, Nucl.
Phys. B796 (2008) 521, arXiv:0709.0398

\bibitem{3E} E.H Saidi, \emph{BPS and non BPS 7D Black Attractors in
M-Theory on K3}, arXiv:0802.0583,\newline E.H Saidi, \emph{On Black
Hole Effective Potential in 6D/7D N=2 Supergravity},
Nucl.Phys.B803:235-276,2008, arXiv:0803.0827,

\bibitem{3F} E.H Saidi, \emph{Computing the Scalar Field Couplings in 6D Supergravity},
Nucl.Phys.B803:323-362,2008, arXiv:0806.3207,

\bibitem{5A} R. Ahl Laamara, L. B. Drissi, F. Z. Hassani, E. H. Saidi, A.
Soumail, \emph{Intersecting Black Attractors in} \emph{8D}
$\mathcal{N}=1$\ \emph{Supergravity}, arXiv:1011.3299, LPHE-MS-10-03
/ CPM-1002, To appear in NPB,

\bibitem{6A} S. Katz, P. Mayr, C. Vafa, \emph{Mirror symmetry and Exact
Solution of 4D }$\mathcal{N}\emph{=2}$\emph{\ Gauge Theories }I,
Adv.Theor.Math.Phys. 1 (1998) 53-114, hep-th/9706110,

\bibitem{6B} M. Ait Ben Haddou, A. Belhaj, E.H. Saidi, \emph{Geometric
Engineering of N=2 CFT}$_{4}$\emph{s,} Nucl.Phys. B674 (2003)
593-614, arXiv:hep-th/0307244,

\bibitem{4A} S. Ferrara, D. Z. Freedman and P. Van Nieuwenhuizen, \emph{%
Progress Toward a Theory of Supergravity}, Phys. Rev. D13 (1976) 3214,%
\newline
S. Deser and B. Zumino, \emph{Consistent Supergravity}, Phys. Lett.
62B (1976) 335.

\bibitem{4B} E. Cremmer, B. Julia and J. Scherk, \emph{Supergravity theory
in 11 dimensions, }Phys.Lett. B 76 (1978) 409,

\bibitem{4C} S. Ferrara, J.G. Taylor, \emph{Supergravity'81}, \emph{%
Proceedings of the 1st School on Supergravity}, International Centre
for Theoretical Physics, Trieste, Italy,

\bibitem{7A} Andrei Micu, \emph{Heterotic type IIA duality with fluxes -
towards the complete story}, arXiv:1009.2357,

\bibitem{7B} Jan Louis, Andrei Micu, \emph{Heterotic-type IIA duality with
fluxes}, JHEP 0703:026,2007, arXiv:hep-th/0608171,

\bibitem{7C} Mitsuko Abe, Masamichi Sato, \emph{Puzzles on the Duality
between Heterotic and Type IIA Strings}, Phys.Lett. B467 (1999)
218-224, arXiv:hep-th/9904155,

\bibitem{8A} F. Cachazo, B. Fiol, K. Intriligator, S. Katz, C. Vafa, \emph{A
Geometric Unification of Dualities}, Nucl.Phys. B628 (2002) 3-78,
hep-th/0110028,

\bibitem{8B} R. Ahl Laamara, M. Ait Ben Haddou, A Belhaj, L.B Drissi, E.H
Saidi, \emph{RG Cascades in Hyperbolic Quiver Gauge Theories,
}Nucl.Phys. B702 (2004) 163-188, arXiv:hep-th/0405222.
\end{thebibliography}
\end{document}